\documentclass{article}

    \PassOptionsToPackage{numbers, compress}{natbib}

\usepackage[preprint]{neurips_2020}

\usepackage[utf8]{inputenc} 
\usepackage[T1]{fontenc}    
\usepackage{hyperref}       
\usepackage{url}            
\usepackage{booktabs}       
\usepackage{nicefrac}       
\usepackage{microtype}      
\usepackage{graphicx}
\usepackage{subfig}
\usepackage[export]{adjustbox}
\usepackage{amsmath,amsfonts,amssymb}

\title{NodeNet: A Graph Regularised Neural Network for Node Classification}

\author{%
    Shrey Dabhi\\
    Robert Bosch Enigneering and Business Solutions, Bengaluru, India\\
    Institute of Technology, Nirma University, Ahmedabad, India\\
    \texttt{16bit039@nirmauni.ac.in}\\
    \And
    Manojkumar Parmar\\
    Robert Bosch Enigneering and Business Solutions, Bengaluru, India\\
    HEC Paris, Jouy-en-Josas Cedex, France\\
    \texttt{manojkumar.parmar@bosch.com}\\
}

\begin{document}

\maketitle

\begin{abstract}
    Real-world events exhibit a high degree of interdependence and connections, and hence data points generated also inherit the linkages. However, the majority of AI/ML techniques leave out the linkages among data points. The recent surge of interest in graph-based AI/ML techniques is aimed to leverage the linkages. Graph-based learning algorithms utilize the data and related information effectively to build superior models. Neural Graph Learning (NGL) is one such technique that utilizes a traditional machine learning algorithm with a modified loss function to leverage the edges in the graph structure. In this paper, we propose a model using NGL - NodeNet, to solve node classification task for citation graphs. We discuss our modifications and their relevance to the task. We further compare our results with the current state of the art and investigate reasons for the superior performance of NodeNet.
\end{abstract}

\section{Introduction}

Many machine learning tasks in the domain of computer vision and natural language processing have been revolutionized by end-to-end deep learning approaches. While deep learning can effectively capture patterns in structure data, there is an increasing number of tasks where representing data in the form of graphs is more intuitive and improves the performance significantly.

\begin{figure}[h]
    \centering
    \subfloat[b][2D Convolution. Every pixel is considered as a node and the number of neighbours is determined by the size of filter. It takes weighted average values of green node and its neighbours.]{\includegraphics[height=0.35\linewidth]{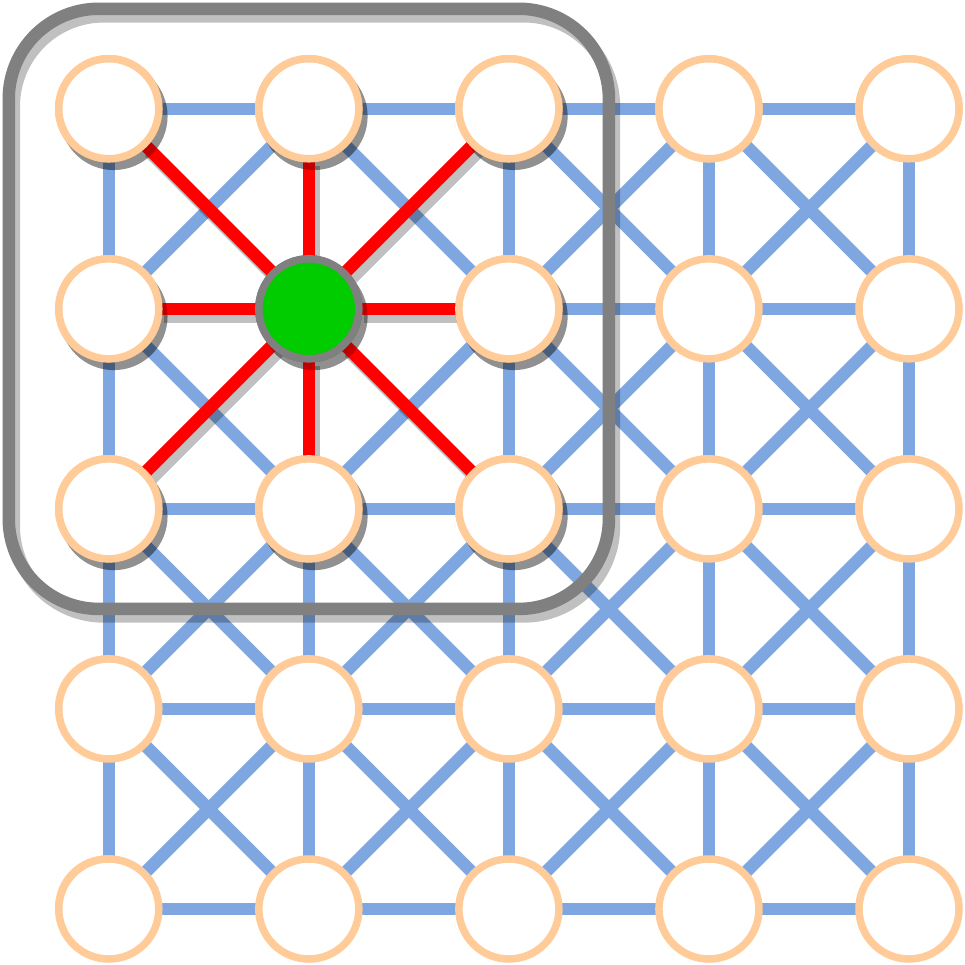}}
    \hspace{0.05\linewidth}
    \subfloat[b][Graph Convolution. A simple operation to obtain the latent representation would be to take weighted average values of green node and its neighbours.]{\includegraphics[height=0.35\linewidth]{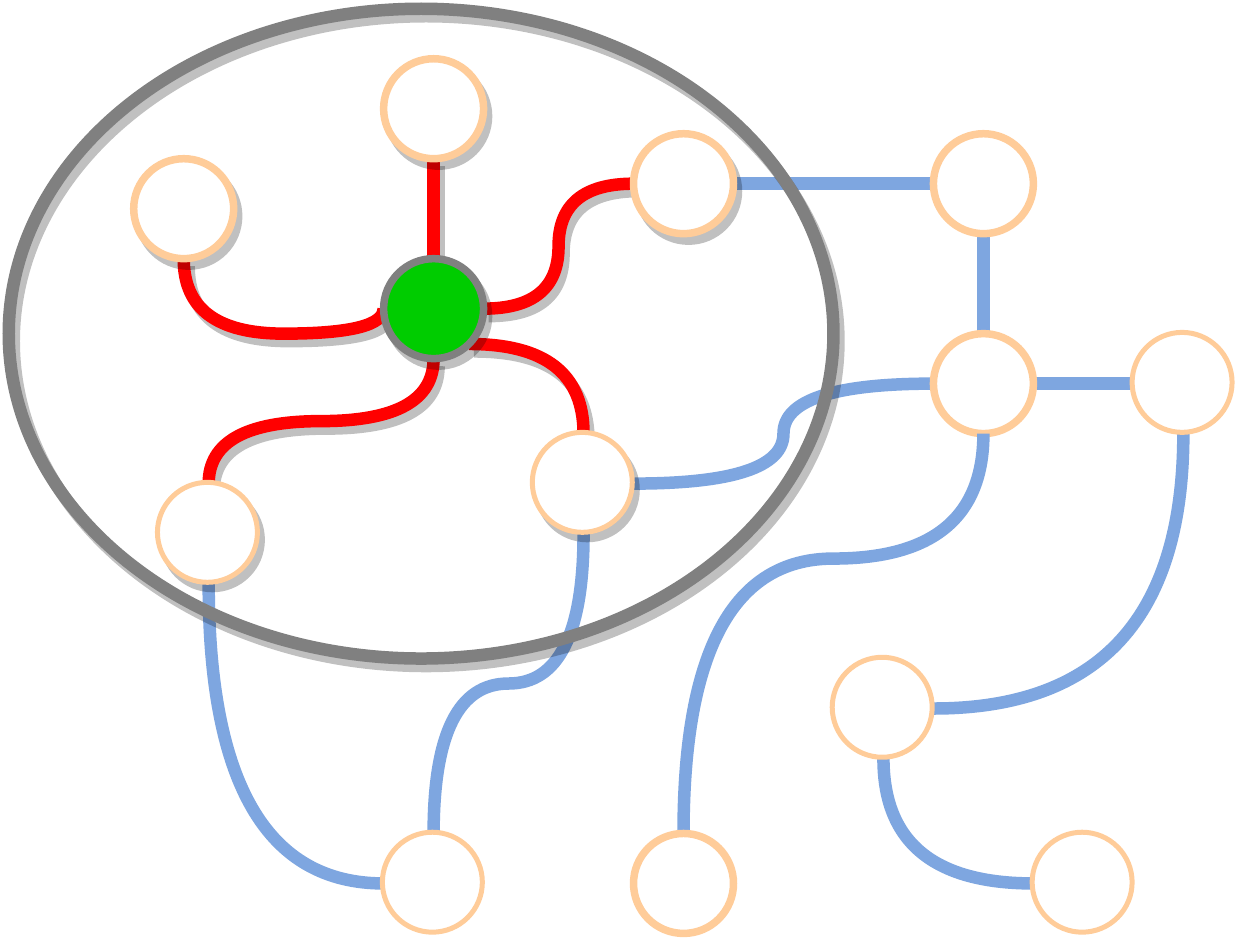}}
    \caption{2D Convolution vs. Graph Convolution (adapted from \cite{9046288})}
    \label{fig:img_as_graph}
\end{figure}

Graphs are relational in nature, in the sense that the connections amongst the nodes denote some form of relationship between them. A graph also provides mechanism to represent multiple relationships in the form of link orientation and edge-level features. More importantly graphs are ubiquitous. Complex molecular structures, citation networks, social networks, computation graphs, interaction graphs, etc. are all examples of graphs which can be used to solve different machine learning tasks more effectively. A graph based learning system can model the complex buying behaviour in a better way than the traditional models to make highly accurate recommendations. In fact as shown in Figure \ref{fig:img_as_graph} images can be considered as just a special case of graphs, where each node not lying on the edge has equal number of neighbours and all the nodes can be arranged in a planar systematic arrangement.

According to the taxonomy proposed by \citet{9046288}, Graph Neural Networks (GNNs) can be classified into 4 groups:
\begin{enumerate}
    \item Recurrent Graph Neural Networks (Graph Convolutions Networks or GCNs)
    \item Convolutional Graph Neural Networks (GRNNs)
    \item Graph Autoencoders
    \item Spatio-temporal Graph Neural Networks (STGNNs)
\end{enumerate}
out of which, and GCN and GRNN are of the greatest interest as they are the most suitable GNN architectures for node classification.

In node classification, given a graph(s) with a set of labelled nodes the task is to learn a model which can accurately predict the label of the unlabelled nodes. The sets of labelled and unlabelled nodes may belong to the same graph or to disjoint graphs from the same dataset.

Representing data as graphs poses some very unique challenges. Graphs have varying sizes and topology. Persisting information about the identity of nodes across multiple batches is difficult. Graphs may contain loops and running exhaustive search is not practically feasible. Though majority of these challenges have been conquered by modern GCNs, they still suffer from over-smoothing and over-fitting.

In the following sections we discuss our approach to node classification and how we arrived at it. In Section 2, we discuss GCNs and GRNNs used in past. We also introduce NGL method, which is our inspiration behind NodeNet. In Section 3, we discuss our modifications to NGL. In Section 4, we present our results and our thoughts on why the modifications are relevant to this task.

\section{Literature Survey}

\begin{figure}[h]
    \centering
    \includegraphics[width=\linewidth]{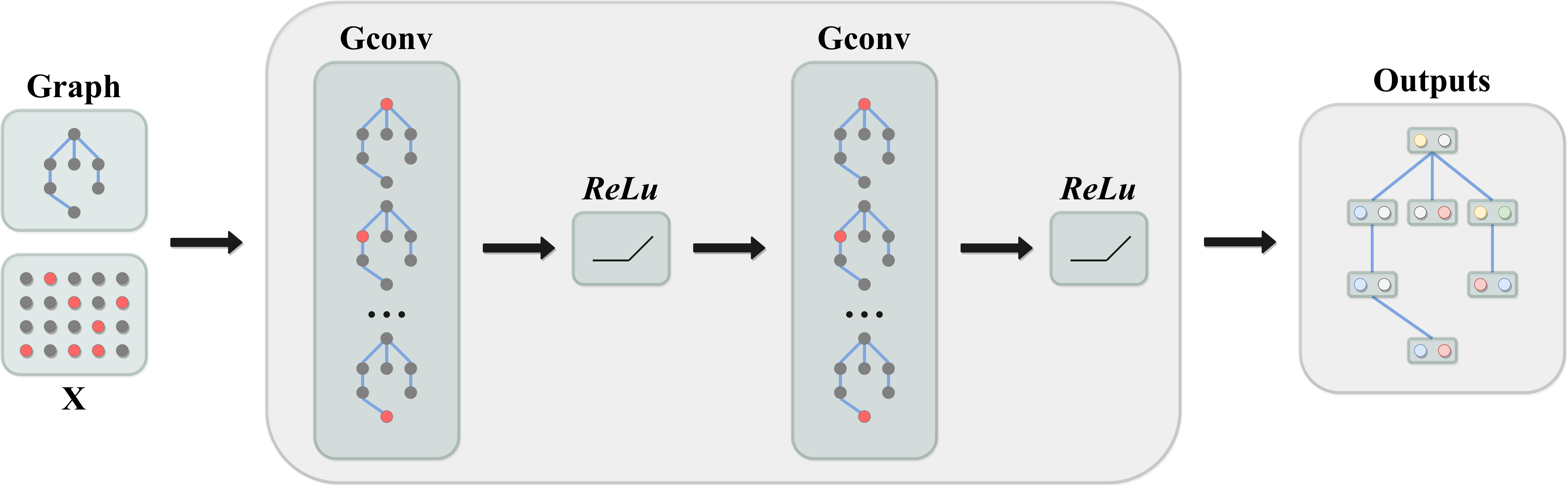}
    \caption{A GCN with multiple graph convolutional layers for node classification (adapted from \cite{9046288})}
    \label{fig:gcn_nc}
\end{figure}

Akin to CNNs, GCNs stack multiple graph convolutional layers to extract high-level node features, followed by fully-connected layers for classification. Figure \ref{fig:gcn_nc} shows a simple GCN with 2 convolutional layers for node classification. These networks are trained as end-to-end models without any separated steps for feature extraction or manipulation

\begin{figure}[h]
    \centering
    \includegraphics[width=\linewidth]{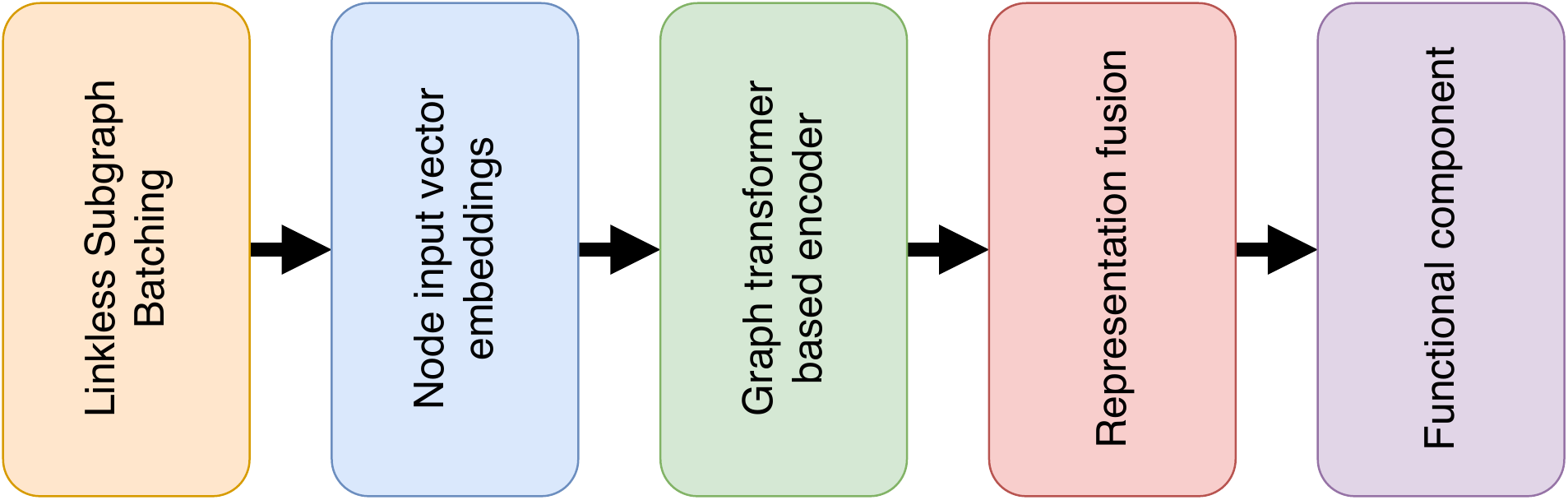}
    \caption{Architecture of the Graph-BERT Model (adapted from \cite{zhang2020graphbert})}
    \label{fig:graph_bert}
\end{figure}

Other approach to node classification, inspired from transformer models used in natural language processing, employ graph-transformer layers and attention models. For example, \citet{zhang2020graphbert} discusses Graph-BERT, inspired from BERT uses statistical and spectral approaches over samples subgraphs to generate input embedding and feed them to a graph-transformer based encoder. The output of the transformer is then used for classification. The complete architecture of Graph-BERT is shown in Figure \ref{fig:graph_bert}. Other approaches like the one proposed by \citet{sun2019adagcn}, follow a RNN like architecture which has proven to be quite good in NLP domain.

These approaches have shown very good results. However, the components have to be designed from scratch for each dataset. The approach proposed by \citet{10.1145/3159652.3159731} is known as Neural Graph Learning (NGL). In NGL, a regularization term is added to the classification loss so that the neural network also learns to associate the label more strongly with certain node features which are related to a particular class label. This regularisation term, Graph Loss, is calculated using a distance metric, node-level features of neighbours and edge-level features. The mathematical formulation of the overall cost function is given in Eqn. \ref{eqn:cost_ngm}.
\begin{align}
\mathcal{C}_{\mathrm{NGL}}(\theta) &= \sum_{n=1}^{V_l} c (g_\theta(x_n), y_n) \nonumber \\
& \quad + \alpha_1 \sum_{(u,v)\in \mathcal{E}_{LL}} w_{uv} d(h_\theta(x_u), h_\theta(x_v)) \nonumber \\ 
& \quad + \alpha_2 \sum_{(u,v)\in \mathcal{E}_{LU}} w_{uv} d(h_\theta(x_u), h_\theta(x_v)) \nonumber \\ 
& \quad + \alpha_3 \sum_{(u,v)\in \mathcal{E}_{UU}} w_{uv} d(h_\theta(x_u), h_\theta(x_v)) \label{eqn:cost_ngm}
\end{align}
where $\mathcal{E}_{LL}$, $\mathcal{E}_{LU}$, and $\mathcal{E}_{UU}$ are sets of labeled-labeled, labeled-unlabeled and unlabeled-unlabeled edges correspondingly, $g(\cdot)$ is the label predicted by the neural network, $h(\cdot)$ represents the latent representations of the inputs produced by the neural network, $d(\cdot)$ is a distance metric, and $\{\alpha_1, \alpha_2, \alpha_3\}$ are hyperparameters. This cost function accounts for the label propogation cost and the neural network cost. \citet{10.1145/3159652.3159731} vouch for using $l$-1 norm (Eqn. \ref{eqn:l1}) or $l$-2 norm (Eqn. \ref{eqn:l2}) distance metric for $d(\cdot)$. We build upon the concept of NGL to propose the architecture of NodeNet.

\begin{gather}
    d(\vec{a},\vec{b}) = \Sigma |a_i - b_i| \label{eqn:l1} \\
    d(\vec{a},\vec{b}) = |\vec{a} - \vec{b}| \label{eqn:l2}
\end{gather}

While NGL does improve the efficiency, it comes with an added benefit that we do not require the inforation about the neighbor nodes at the time of inference. In our opinion it could be because the neural network itself memorizes information about the edges in the graph via the loss function.

In this paper we consider 3 very popular citation network datasets to validate our modifications:  Cora, Citeseer, and Pubmed. In citation networks, nodes correspond to scientific documents, edges correspond to citation links, and each node has a feature vector as well as a class label. The statistics of the 3 datasets are given in Table \ref{tab:data}.

\begin{table}[h]
    \caption{Dataset statistics}
    \label{tab:data}
    \begin{tabular*}{\linewidth}{c @{\extracolsep{\fill}} ccccc}
        \toprule
        Dataset & \# nodes & \# edges & \# features & \# classes & Type of feature \\ \midrule
        Cora \cite{McCallum_2000} & 2708 & 5429 & 1433 & 7 & Binary Count Vector \\
        Citeseer \cite{10.1145/276675.276685} & 3327 & 4732 & 3703 & 6 & Binary Count Vector \\
        Pubmed \cite{Sen_Namata_Bilgic_Getoor_Galligher_Eliassi-Rad_2008} & 19717 & 44338 & 500 & 3 & TF-IDF Vector \\ \bottomrule
    \end{tabular*}
\end{table}

We felt that even though the current state-of-the-art techniques provide great results, they do not satisfactorily address the problems of over-fitting and over-smoothing. Majority of the GNNs designed for node classification do not contain more than 2 layers as the effect of over-smoothing is very high in deeper neural networks. From an application point-of-view, the whole graph or a sub-graph may or may not be available at the time of inference, which make conventional GNNs unsuitable for these kind of applications. NGL \cite{10.1145/3159652.3159731} addresses these issues to a great extent, but falls short in terms of accuracy. With NodeNet we aim to cross this barrier of accuracy as well

\section{Proposed Method}

 When we came across the approach of \citet{10.1145/3159652.3159731}, we saw an opportunity to leverage a large knowledge base of neural network architectures and combine it with the extra information extracted from the edges of the graph. We propose changes in 3 areas, pre-processing, network architecture and regularization term, to arrive at NodeNet.

\subsection{Pre-processing}

Generally term frequency–inverse document frequency (TF-IDF) \cite{10.5555/553876} vectors perform better in any NLP task as compared to binary count vectors. This applies to classification as well. The words and absolute word counts are not available for the documents / nodes in Cora and CiteSeer datasets. Hence, we prepare a modified TF-IDF vector using Eqn. \ref{eqn:tfidf}.
\begin{gather}
    mTFIDF_{i,j} = \frac{IDF_{i,j}}{n_i} \label{eqn:tfidf} \\
    IDF_{i,j} = log\left(\frac{N}{1 + N_{t_{i,j}}}\right) + 1 \label{eqn:idf}
\end{gather}
In Eqn. \ref{eqn:tfidf} and Eqn. \ref{eqn:idf}, the values are being calculated for $j^{th}$ term of $i^{th}$ document. Additionally, $n_i$ stands for total number of terms in $i^{th}$ document, $N$ stands for the total number of documents and $N_{t_{i,j}}$ stands for number of documents in which the term $t_{i,j}$ appears. Eqn. \ref{eqn:tfidf} is the modified TF-IDF \cite{manning_raghavan_schutze_2008} score and Eqn. \ref{eqn:idf} is the smooth inverse document frequency weight \cite{manning_raghavan_schutze_2008}.

\subsection{Network Architecture}

\citet{10.1145/3159652.3159731} propose a network architecture containing only fully-connected (dense) and activation layers for node classification on PubMed dataset. We observed that this network easily overfits the dataset. Batch Normalization is known to have a regularization effect while enabling training with a relatively higher learning rate \cite{ioffe2015batch} and Dropout \cite{9046288} is a very effective way to combat over fitting. We first added the batch normalization layer, but were still able to observe effect of over fitting. Hence we also added the dropout layer.

\begin{figure}[ht]
    \centering
    \includegraphics[width=\linewidth]{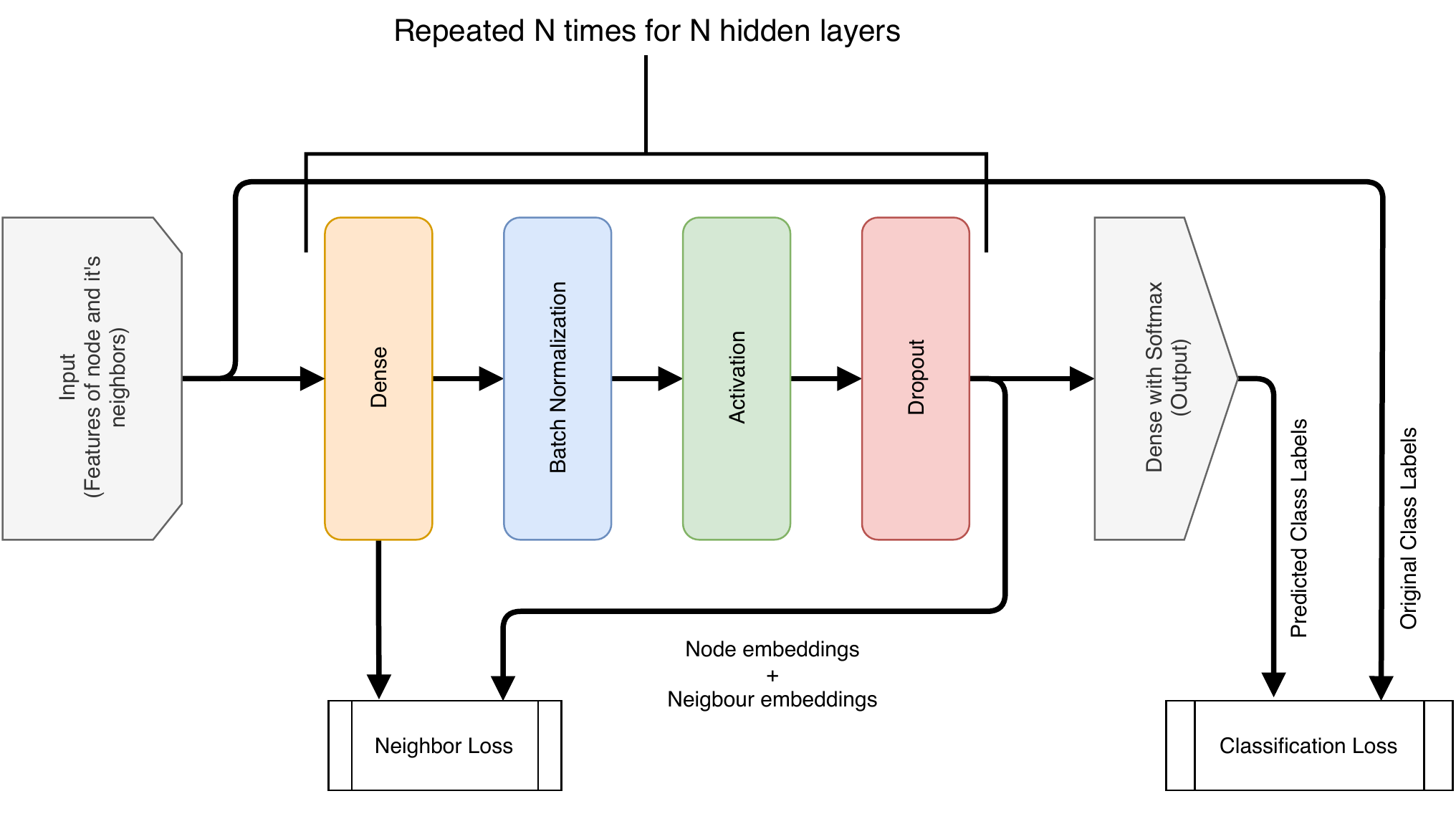}
    \caption{Network Architecture}
    \label{fig:network_architecture}
\end{figure}

\subsection{Regularization term}

We changed the distance metric from $l$-2 norm to cosine similarity. The reason being that it better represents the degree of similarity between 2 documents \cite{Singhal01moderninformation}. Such a similarity can only be found along a citation edge as a paper would only cite other papers which are relevant to another, and hence there will be more similarity in the unique words used.

\begin{equation}
    cosineSim(\vec{a},\vec{b}) = \frac{\vec{a} \cdot \vec{b}}{|\vec{a}| \times |\vec{b}|}
    \label{eqn:cosine}
\end{equation}
where $cosineSim(\vec{a},\vec{b})$ function calculates the cosine similarity between 2 vectors, $\vec{a}$ and $\vec{b}$.

By replacing $d(\cdot)$ in Eqn. \ref{eqn:cost_ngm} by the cosine similarity function, we get the following final cost function for NodeNet:
\begin{align}
\mathcal{C}_{\mathrm{NodeNet}}(\theta) &= \sum_{n=1}^{V_l} c (g_\theta(x_n), y_n) \nonumber \\
& \quad + \alpha_1 \sum_{(u,v)\in \mathcal{E}_{LL}} w_{uv} cosineSim(h_\theta(x_u), h_\theta(x_v)) \nonumber \\ 
& \quad + \alpha_2 \sum_{(u,v)\in \mathcal{E}_{LU}} w_{uv} cosineSim(h_\theta(x_u), h_\theta(x_v)) \nonumber \\ 
& \quad + \alpha_3 \sum_{(u,v)\in \mathcal{E}_{UU}} w_{uv} cosineSim(h_\theta(x_u), h_\theta(x_v)) \label{eqn:cost_nodenet}
\end{align}

\section{Results}

As can be seen from the results presented in Table \ref{tab:res}, NodeNet surpasses the current state-of-the-art for all the 3 data sets. We have tried to tackle the issues of over-smoothing and over-fitting by using appropriate regularization techniques.

\begin{figure}[h]
    \centering
    \subfloat[b][PubMed]{\includegraphics[width=0.32\linewidth,frame]{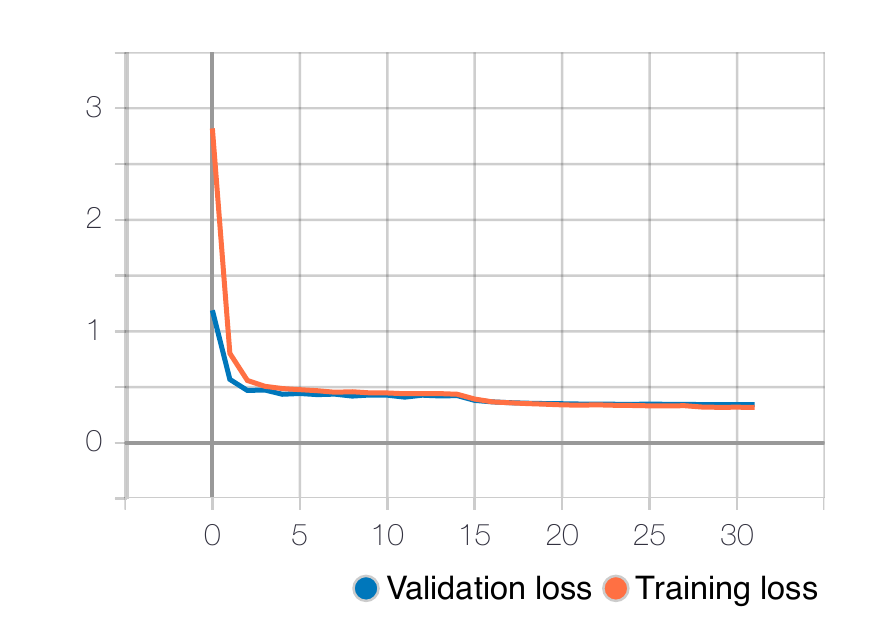}}
    \hspace{0.01\linewidth}
    \subfloat[b][Cora]{\includegraphics[width=0.32\linewidth,frame]{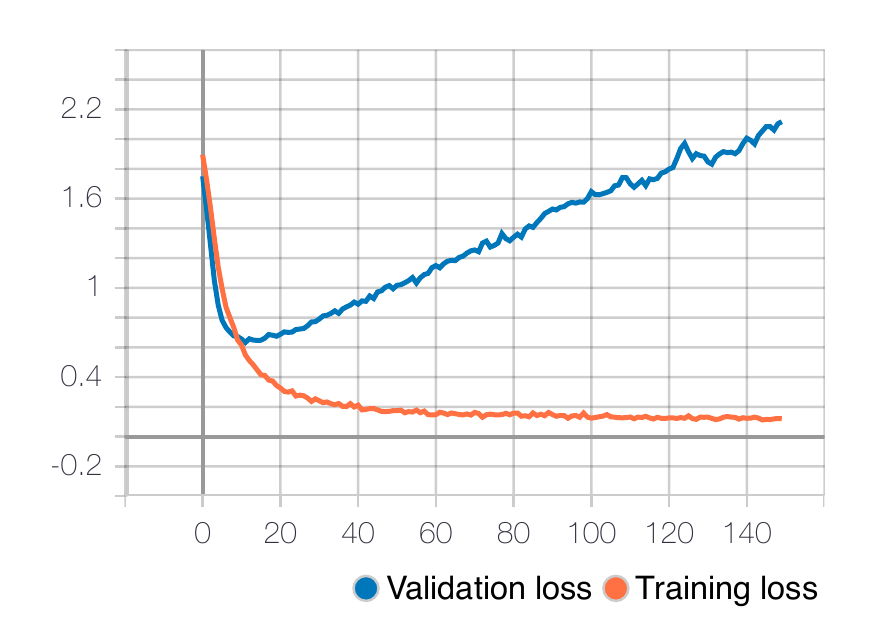}}
    \hspace{0.01\linewidth}
    \subfloat[b][Cora using TF-IDF vectors]{\includegraphics[width=0.32\linewidth,frame]{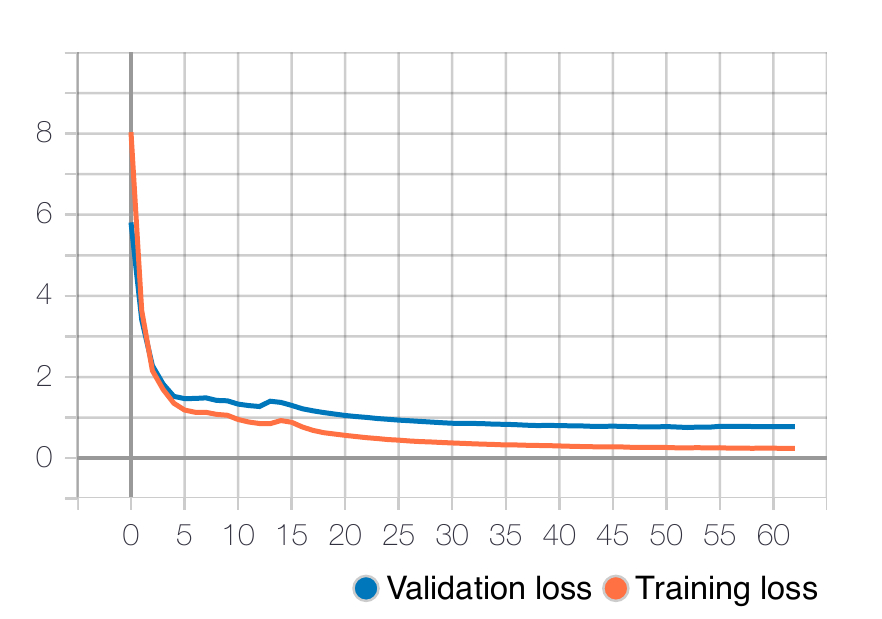}}
    \caption{Loss curves of NodeNet for different datasets}
    \label{fig:loss}
\end{figure}

We observed from the loss curve for PubMed data set (Fig. \ref{fig:loss}(a)) that the neural networks has a faster and more stable convergence over the PubMed dataset which has TF-IDF vectors as compared to Cora (Fig. \ref{fig:loss}(b)) and CiteSeer data sets which have binary count vectors. Hence, we decided to transform binary count vectors into modified TF-IDF vectors. This improved the learning ability and stabilized the convergence (Fig. \ref{fig:loss}(c)), at the cost of a small decline in the accuracy.

\begin{table}[h]
    \caption[Comparison of accuracy]{Comparison of accuracy against state-of-the-art algorithms \footnotemark}
    \label{tab:res}
    \begin{tabular*}{\linewidth}{c @{\extracolsep{\fill}} cc}
        \toprule
        Dataset & State-of-the-art Algorithm & NodeNet \\ \midrule
        Cora & 86.00 (DFNet-ATT \cite{NIPS2019_8834}) & 86.80 \\
        Cora with additional training data & 89.48 (SplineCNN \cite{Fey_2018_CVPR}) & - \\
        Cora with TF-IDF vectors & - & 85.17 \\
        Citeseer & 78.70 (GCN-LPA \cite{wang2020unifying}) & 80.09 \\
        Citeseer with TF-IDF vectors & - & 78.02 \\
        PubMed & 87.80 (GCN-LPA \cite{wang2020unifying}) & 90.21 \\ \bottomrule
    \end{tabular*}
\end{table}
\footnotetext{As of May 31, 2020 on \href{https://paperswithcode.com/task/node-classification}{Papers With Code}}

\section{Summary}

Using Neural Graph Learning we can leverage the advantages provided by graphs and the vast knowledge-base of neural network architectures. NodeNet is the outcome of coming together of years of knowledge from NLP domain and the advantages of representing data in form of a graph. We demonstrates in this paper that NGL can be further modified to improve it's performance for node classification. Proposed NodeNet modifies pre processing step, architecture and loss function to achieve superior accuracy results while avoiding over-fitting. The NodeNet can be further adapted to make it suitable for generic node classification task.

\section*{Broader Impact}
Research communities and industries building applications based on node classification or working on data which can be modelled as graphs may benefit from this work. No one is put at disadvantage. If the system fails it will provide incorrect labels. If the system is integrated with other systems downstream then the evaluation of impact of failure has to be done separately. If decision are to be taken based on the output, then the impact has to be evaluated from an action-oriented perspective. We do not believe that the proposed algorithm leverages any biases in the data.

\begin{ack}
We want to thank Sri Krishnan V, Mohan B V and Adit Shah from Robert Bosch Engineering and Business Solutions Private Limited, India, for their valuable comments, contributions and continued support to the project. We are grateful to all experts for providing us with their valuable insights and informed opinions ensuring completeness of our study.
\end{ack}

\bibliography{references}

\begin{thebibliography}{14}
\providecommand{\natexlab}[1]{#1}
\providecommand{\url}[1]{\texttt{#1}}
\expandafter\ifx\csname urlstyle\endcsname\relax
  \providecommand{\doi}[1]{doi: #1}\else
  \providecommand{\doi}{doi: \begingroup \urlstyle{rm}\Url}\fi

\bibitem[{Wu} et~al.(2020){Wu}, {Pan}, {Chen}, {Long}, {Zhang}, and
  {Yu}]{9046288}
Z.~{Wu}, S.~{Pan}, F.~{Chen}, G.~{Long}, C.~{Zhang}, and P.~S. {Yu}.
\newblock A comprehensive survey on graph neural networks.
\newblock \emph{IEEE Transactions on Neural Networks and Learning Systems},
  pages 1--21, 2020.

\bibitem[Zhang et~al.(2020)Zhang, Zhang, Xia, and Sun]{zhang2020graphbert}
Jiawei Zhang, Haopeng Zhang, Congying Xia, and Li~Sun.
\newblock Graph-bert: Only attention is needed for learning graph
  representations, 2020.

\bibitem[Sun et~al.(2019)Sun, Lin, and Zhu]{sun2019adagcn}
Ke~Sun, Zhouchen Lin, and Zhanxing Zhu.
\newblock Adagcn: Adaboosting graph convolutional networks into deep models,
  2019.

\bibitem[Bui et~al.(2018)Bui, Ravi, and Ramavajjala]{10.1145/3159652.3159731}
Thang~D. Bui, Sujith Ravi, and Vivek Ramavajjala.
\newblock Neural graph learning: Training neural networks using graphs.
\newblock In \emph{Proceedings of the Eleventh ACM International Conference on
  Web Search and Data Mining}, WSDM ’18, page 64–71, New York, NY, USA,
  2018. Association for Computing Machinery.
\newblock ISBN 9781450355810.
\newblock \doi{10.1145/3159652.3159731}.
\newblock URL \url{https://doi.org/10.1145/3159652.3159731}.

\bibitem[McCallum et~al.(2000)McCallum, Nigam, Rennie, and
  Seymore]{McCallum_2000}
Andrew~Kachites McCallum, Kamal Nigam, Jason Rennie, and Kristie Seymore.
\newblock Automating the construction of internet portals with machine
  learning.
\newblock \emph{Information Retrieval}, 3\penalty0 (2):\penalty0 127--163,
  2000.
\newblock \doi{10.1023/a:1009953814988}.
\newblock URL \url{https://doi.org/10.1023%2Fa%3A1009953814988}.

\bibitem[Giles et~al.(1998)Giles, Bollacker, and
  Lawrence]{10.1145/276675.276685}
C.~Lee Giles, Kurt~D. Bollacker, and Steve Lawrence.
\newblock Citeseer: An automatic citation indexing system.
\newblock In \emph{Proceedings of the Third ACM Conference on Digital
  Libraries}, DL ’98, page 89–98, New York, NY, USA, 1998. Association for
  Computing Machinery.
\newblock ISBN 0897919653.
\newblock \doi{10.1145/276675.276685}.
\newblock URL \url{https://doi.org/10.1145/276675.276685}.

\bibitem[Sen et~al.(2008)Sen, Namata, Bilgic, Getoor, Galligher, and
  Eliassi-Rad]{Sen_Namata_Bilgic_Getoor_Galligher_Eliassi-Rad_2008}
Prithviraj Sen, Galileo Namata, Mustafa Bilgic, Lise Getoor, Brian Galligher,
  and Tina Eliassi-Rad.
\newblock Collective classification in network data.
\newblock \emph{AI Magazine}, 29\penalty0 (3):\penalty0 93, Sep. 2008.
\newblock \doi{10.1609/aimag.v29i3.2157}.
\newblock URL
  \url{https://www.aaai.org/ojs/index.php/aimagazine/article/view/2157}.

\bibitem[Baeza-Yates and Ribeiro-Neto(1999)]{10.5555/553876}
Ricardo~A. Baeza-Yates and Berthier Ribeiro-Neto.
\newblock \emph{Modern Information Retrieval}.
\newblock Addison-Wesley Longman Publishing Co., Inc., USA, 1999.
\newblock ISBN 020139829X.

\bibitem[Manning et~al.(2008)Manning, Raghavan, and
  Schütze]{manning_raghavan_schutze_2008}
Christopher~D. Manning, Prabhakar Raghavan, and Hinrich Schütze.
\newblock \emph{Introduction to Information Retrieval}, page 100–123.
\newblock Cambridge University Press, USA, 2008.
\newblock ISBN 0521865719.
\newblock \doi{10.1017/CBO9780511809071.007}.

\bibitem[Ioffe and Szegedy(2015)]{ioffe2015batch}
Sergey Ioffe and Christian Szegedy.
\newblock Batch normalization: Accelerating deep network training by reducing
  internal covariate shift, 2015.

\bibitem[Singhal(2001)]{Singhal01moderninformation}
Amit Singhal.
\newblock Modern information retrieval: a brief overview.
\newblock \emph{BULLETIN OF THE IEEE COMPUTER SOCIETY TECHNICAL COMMITTEE ON
  DATA ENGINEERING}, 24:\penalty0 35--43, 2001.

\bibitem[Wijesinghe and Wang(2019)]{NIPS2019_8834}
W.~O. K. Asiri~Suranga Wijesinghe and Qing Wang.
\newblock Dfnets: Spectral cnns for graphs with feedback-looped filters.
\newblock In \emph{Advances in Neural Information Processing Systems 32}, pages
  6009--6020. Curran Associates, Inc., 2019.
\newblock URL
  \url{http://papers.nips.cc/paper/8834-dfnets-spectral-cnns-for-graphs-with-feedback-looped-filters.pdf}.

\bibitem[Fey et~al.(2018)Fey, Eric~Lenssen, Weichert, and
  Müller]{Fey_2018_CVPR}
Matthias Fey, Jan Eric~Lenssen, Frank Weichert, and Heinrich Müller.
\newblock Splinecnn: Fast geometric deep learning with continuous b-spline
  kernels.
\newblock In \emph{The IEEE Conference on Computer Vision and Pattern
  Recognition (CVPR)}, June 2018.

\bibitem[Wang and Leskovec(2020)]{wang2020unifying}
Hongwei Wang and Jure Leskovec.
\newblock Unifying graph convolutional neural networks and label propagation,
  2020.

\end{thebibliography}
\bibliographystyle{unsrtnat}

\end{document}